\documentstyle[aps,epsfig,a5,tighten,psfig,epsfig]{revtex} 
\newcommand{\ga}{\alpha}
\newcommand{\gb}{\beta}
\newcommand{\gc}{\gamma}
\newcommand{\gd}{\delta}
\newcommand{\gk}{\kappa}

\newcommand{\gs}{\sigma}

\begin{document}
\title{Few-Body Correlations in Fermi Systems}
\author{M. Beyer}
\address{FB Physik, University of Rostock, 18051 Rostock, Germany}
\maketitle

\vspace{1ex}
Interacting quantum systems with strong or long-range
interactions exhibit quite a rich phase structure.  Cluster formation
and superconductivity are examples.  These phenomena are also expected
in the astrophysical context, e.g., during the formation or in the
structure of neutron stars. To describe these phenomena a proper
treatment has to go beyond the simple picture of noninteracting
quasiparticles. An appealing formalism for a systematic approach is
provided by the framework of Dyson equations.  Within an equal
(imaginary) time formalism Dyson equations can be derived for an
arbitrary large cluster embedded in a medium~\cite{duk98}. For
practical use and the sake of simplicity the medium is treated as
uncorrelated to derive the respective $n$-body cluster Green
functions. Further, we neglect ``backward'' propagating particles, so
the Fock spaces for different number of particles $n$ are
disconnected. This way it is possible to derive effective in-medium
$n$-body equations that can be solved rigorously with few-body
techniques~\cite{Beyer:1996rx,Beyer:1997sf,Beyer:1999tm,Beyer:1999zx,Kuhrts:2000jz,Beyer:1999xv,Beyer:2000ds,Kuhrts:2000zs,alt67,san74,alt72,san75}.

These resulting two-, three-, and four-body equations elaborated here
include the dominant medium effects in a systematic way. These are the
self energy corrections for masses and the Pauli blocking that in turn
leads to a change of binding energies, viz. change of the masses of
clusters, and change of reaction rates.  Further, within this approach
the critical temperatures for condensation (of bosons containing two or
four particles) are calculated.

\begin{figure}[t]
\begin{center}
\epsfig{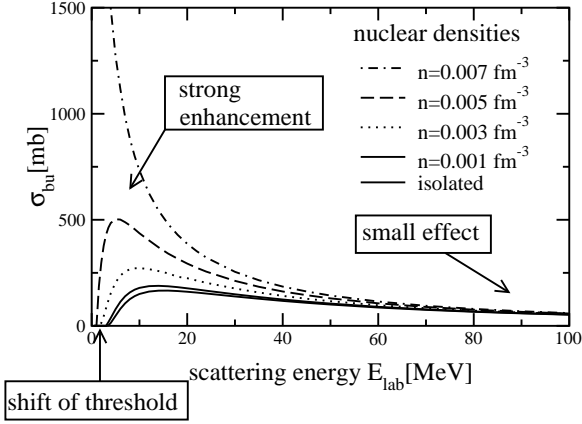}
\caption{\label{fig:Across} 
    Break-up cross section for different densities of nuclear matter for temperature $T=10$ MeV}
\end{center}
\end{figure}

\begin{figure}[t]
\begin{center}
\psfig{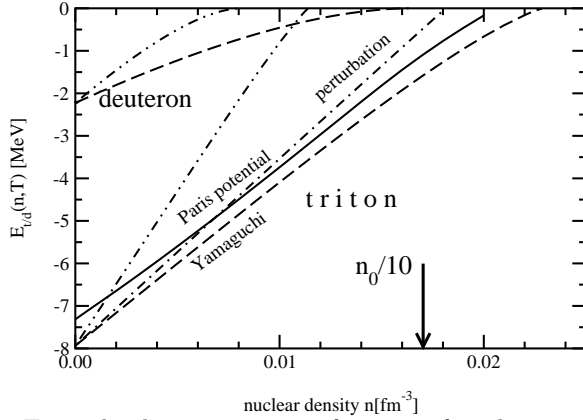}
\caption{\label{fig:Atri0} 
  Triton binding energy as a function of nuclear matter density.
  Dashed-dot-dot $T=10$ MeV, other $T=20$ MeV.}
\end{center}
\end{figure}

Defining $H_0=\sum_{i=1}^n \varepsilon_i$ with the quasi-particle self
energy 
\begin{equation}
\varepsilon_1 = k^2_1/2m_1+\sum_{2}V_2(12,\widetilde{12})f_2
\end{equation}
 and the Fermi function $f_1\equiv f(\varepsilon_1) =
1/(e^{\gb(\varepsilon_1 - \mu)}+1),$ the $n$-particle cluster
resolvent $G_0$ is
\begin{equation}
G_0(z) = (z-  H_0)^{-1}
\;{N} \equiv R_0(z)\;{ N}.
\end{equation}
Here $G_0$, $H_0$, and $N$ are matrices in $n$ particle space and $z$ denotes the Matsubara frequency~\cite{fet71}. The
Pauli-blocking factors for $n$-particles are
\begin{equation}
N=\bar f_1\bar f_2 \dots \bar f_n
\pm f_1f_2\dots f_n,\qquad\bar f=1-f
\end{equation}
{Note: $NR_0=R_0N$.}  Defining the effective potential $V\equiv
\sum_{\mathrm{pairs}\;\ga} { N_2^{\ga}}V_2^{\ga}$ the full
and the channel resolvents are
\begin{eqnarray}
G(z)&=&(z-H_0- V)^{-1}{N}
\equiv R(z){N},\\
G_\ga(z)&=&(z-H_0- { N_2^{\ga}}V_2^{\ga})^{-1}{ N}
\equiv R_\ga(z){ N},
\end{eqnarray}
Note that $V^\dagger\neq V$ and $R(z)N\neq NR(z)$.
For the scattering problem it is convenient to define the in-medium
AGS operator $U_{\gb\ga}(z)$
\begin{equation}
R(z)=\gd_{\ga\gb}R_\gb(z) + R_\gb(z) { U_{\gb\ga}(z)} R_\ga(z)
\end{equation}
that after some algebra leads to the in-medium AGS equation 
\begin{equation}
 U_{\gb\ga}(z)=\bar\gd_{\gb\ga}R_0(z)^{-1}+\sum_\gc
\bar\gd_{\gb\gc} 
N_2^\gc T_2^\gc(z) R_0(z) U_{\gc\ga}(z),
\end{equation}
where $\bar\gd_{\gb\ga}=1-\gd_{\gb\ga}$.  The square of this
$t$-operator is directly linked to the differential cross section for
the scattering process $\ga\rightarrow\gb$. The driving kernel
consists
 of the two-body $t$-matrix derived in the same
formalism, however given earlier and known as Feynman-Galitskii
equation\cite{fet71}
\begin{eqnarray}\nonumber
T_2^\gamma(z) &=&   V_2^\gamma + 
 V_2^\gamma { N^\gc_2}R_0(z)  T_2^\gamma(z)\nonumber.
\end{eqnarray}
A numerical solution using a coupled Yamaguchi potential has been
given in Ref.~\cite{Beyer:1996rx}. For a temperature $T=10$ MeV and the
three-body system at rest in the medium results a given in
Fig.\ref{fig:Across}
For the bound state problem it is convenient to introduce form
factors
\begin{equation}
|F_\gb\rangle=\sum_\gc\bar\gd_{\gb\gc} { N_2^\gc}  V_2^\gc 
|\psi_{B_3}\rangle.
\end{equation}
Since the potential is nonsymmetric right and left eigenvectors are
different, although the bound state energies are the same,
\begin{eqnarray}
|F_\ga\rangle
&=&\sum_\gb \bar\gd_{\ga\gb}  
{ N_2^\gb} T_2^\gb(B_3) R_{0}(B_3)|F_\gb\rangle,\\
|\tilde F_\ga\rangle
&=&\sum_\gb \bar\gd_{\ga\gb} T_2^\gb(B_3) { N_2^\gb} R_{0}(B_3)
|\tilde F_\gb\rangle.
\end{eqnarray}
The binding energy depends on $\mu,\;T,\;P_{\mathrm c.m.}$. Results
for $P_{\mathrm c.m.}=0$ are given in Fig.\ref{fig:Atri0} for
different potentials and temperatures. Note that the dependence on
density is rather similar for two different potentials studied,
although the binding energies for the isolated triton differs by 10\%
the Mott density is practically at the same place once the binding
energies are renormalized to each other. For helium the Mott density
is smaller due to the Coulomb force, however for asymmetric nuclear
matter, e.g. $N_p/N_n\simeq 0.72$ (for the $^{129}$Xe$+~^{119}$Sn
reaction) this effect is compensated~\cite{mat00}. The dependence of
the Mott effect on the momentum is given in Ref.~\cite{Beyer:1999zx}.

We now turn to the four-body problem in matter. In addition to having
different channels as for the three body system now the channels
appear in different partitions that makes the four-body problem even
more involved. The partitions of the four-body clusters are denoted by
$\rho,\tau,\sigma,\dots$, e.g., $\rho=(123)(4),(234)(1), \dots$ for
$3+1$-type partitions, or $\rho=(12)(43), (23)(41), \dots$ for
$2+2$-type partitions.  The two-body sub-channels are denoted by pair
indices $\ga,\gb,\gc,\dots$, e.g. pairs $(12)$, $(24)$,\dots The two-
and three-body $t$-matrices have to be defined with respect to the
partitions that leads to additional indices. 
The four-body in-medium homogeneous AGS equation are defined for the
form factors
\begin{equation}
        |{\cal F}_\gb^\gs\rangle = \sum_\tau \bar\gd_{\gs\tau}
        \sum_\ga \bar\gd^\tau_{\gb\ga} R_0^{-1}(B_4) |\psi_\ga\rangle,
\end{equation}
where $\bar\gd^\tau_{\gb\ga}=\bar\gd_{\gb\ga}$, if
$\gb,\ga\subset\tau$ and $\bar\gd^\rho_{\gb\ga}=0$ otherwise and
$|\psi_\ga\rangle$ is the $\alpha$-üparticle wave function. They
read~\cite{Beyer:2000ds}
\begin{equation}
|{\cal F}^\gs_\gb\rangle=\sum_{\tau\gc} \bar\gd_{\gs\tau}
U^\tau_{\gb\gc}(B_4)  R_0(B_4) { N_2^\gc} 
T_2^\gc(B_4) 
R_0(B_4) |{\cal F}^\tau_\gc\rangle,
\end{equation}
where $\ga\subset\gs,\gc\subset\tau$.  A numerical solution of this
equation is rather complex. In order to reduce computational time we
introduce a energy dependent pole expansion (EDPE) that has been
proven useful in many application involving the $\alpha$-particle and
is accurate enough for the present purpose.  However, we have to
generalize the original version of the EDPE because of different right
and left eigenvectors. Details will be omitted here,
see~\cite{Beyer:2000ds}

In the {\em two-body} sub-system the EDPE reads
\begin{eqnarray}
T_\gc(z) &\simeq &\sum_n
|\tilde\Gamma_{\gc n}(z)\rangle t_{\gc n}(z)\langle \Gamma_{\gc n}(z)|
\simeq\sum_n |\tilde g_{\gc n}\rangle t_{\gc n}(z)\langle g_{\gc n}|
\nonumber\\
&=&\sum_n
{  N_2^\gc} |g_{\gc n}\rangle t_{\gc n}(z)\langle g_{\gc n}|.
\label{eqn:Tsep2}
\end{eqnarray}
where we have chosen a Yamaguchi ansatz for the form factors for
simplicity.  Inserting this ansatz into the Feynman-Galitskii equation
determines the propagator $t_{\gc n}(z)$. In the {\em three-body}
sub-system the EDPE expansion reads
\begin{equation}
\langle g_{\gb m}(z)| R_0(z) U^\tau_{\gb\gc}(z)R_0(z)| 
\tilde g_{\gc n}(z)\rangle
\simeq \sum_{t,\mu\nu} |\tilde\Gamma^{\tau t, \mu}_{\gb m}(z)\rangle
t^{\tau t}_{\mu\nu}(z)\langle \Gamma^{\tau t, \nu}_{\gc n}(z)|.
\label{eqn:pole3}
\end{equation}
with the three-body EDPE functions 
\begin{equation}
|\tilde\Gamma^{\tau t, \mu}_{\gb m}(z)\rangle
= \langle g_{\ga n}|R_0(z)| \tilde g_{\gb m}\rangle
t_{\gb m}(B_3) |\tilde\Gamma^{\tau t, \mu}_{\gb m}\rangle,
\end{equation}
that we get from solving the proper Sturmian equations
\begin{eqnarray}
\eta_{t,\mu}|\tilde\Gamma^{\tau t, \mu}_{\ga n}\rangle&=&
\sum_{\gb m}
 \langle g_{\ga n}|R_0(B_3)| \tilde g_{\gb m}\rangle
t_{\gb m}(B_3)|\tilde\Gamma^{\tau t, \mu}_{\gb m}\rangle
\label{eqn:sturm}\\
\eta_{t,\mu}|\Gamma^{\tau t, \mu}_{\ga n}\rangle&=&
\sum_{\gb m}
 \langle \tilde g_{\ga n}|R_0(B_3)|  g_{\gb m}\rangle
t_{\gb m}(B_3)|\Gamma^{\tau t, \mu}_{\gb m}\rangle
\end{eqnarray}
Inserting everything into the homogeneous AGS equations allows us to
redefine the form factors that are now operators in the coordinates of
the $2+2$ or $3+1$ system, respectively
\begin{equation} 
|\Gamma^{\gs s}_\mu\rangle 
= \sum_{\gb m}\langle\Gamma^{\gs s}_{\gb m,\nu}(B_4)|t_{\gb m}(B_4)
\langle g_{\gb m}(B_4)| R_0(B_4)|{\cal F}^\gs_\gb\rangle
\end{equation} 
and therefore the final homogeneous equation
\begin{equation}
|\Gamma^{\gs s}_\mu\rangle =
\sum_{\tau t}\sum_{\nu\gk} \sum_{\gc n} \bar\gd_{\gs\tau}
\langle\Gamma^{\gs s, \nu}_{\gc n}(B_4)|t_{\gc n}(B_4)
|\tilde\Gamma^{\gs s, \mu}_{\gc n}(B_4)\rangle\;
t^{\tau t}_{\mu\gk}(B_4)\;|\Gamma^{\tau t}_\gk\rangle,
\label{eqn:coup4}
\end{equation}
is an effective one-body equation with
and effective potential ${\cal V}$ and an effective resolvent
${\cal G}_0$:
\begin{eqnarray}
{\cal V}^{\gs s,\tau t}_{\mu\nu}(z)
&= &\sum_{\gc n} \bar\gd_{\gs\tau}
\langle\Gamma^{\gs s, \mu}_{\gc n}(z)|t_{\gc n}(z)
|\tilde\Gamma^{\gs s, \nu}_{\gc n}(z)\rangle,
\label{eqn:pot4}\\
{\cal G}^{\gs s,\tau t}_{\mu\nu,0}(z)&=& t^{\tau t}_{\mu\nu}(z).
\end{eqnarray}

\begin{figure}[t]
\begin{center}
\psfig{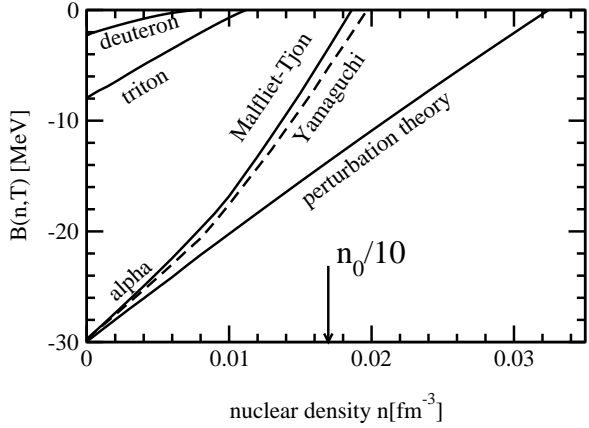}
\caption{\label{fig:amottC} Energy dependence of the binding energy of the
  $\alpha$-particle}
\end{center}
\end{figure}
The binding energies of the two-, three-, and four-body systems are
shown in Fig.\ref{fig:amottC} for a temperature of $T=10$ MeV and a
c.m. momentum of the respective cluster of $P_{\mathrm c.m.}=0$. The
$B=0$ line reflects the respective continuum. Investigating the zeros
of the two-body Joost function the quasi deuteron survives as an
anti-bound state (not resonance) with increasing densities, viz. for
energies above the continuum\cite{bey01}. The fate of the triton and
of the $\alpha$ particle for $B>0$ still needs to investigated as well
as a possible appearance of Efimov states related to $B\rightarrow 0$
of the sub-system. Since the Efimov states are 'excited' states, e.g.
for the three-body system close to the $2+1$ threshold, their blocking
may be smaller since the wave functions contain higher momentum
components. 
Note that the slope of the binding energies as a function of densities
for the larger clusters is also larger. This is a clear indication
that the masses of the clusters change with increasing density. The
Mott density of the $\alpha$ particle appears at
$$
B_4(n_{\mathrm Mott}=0.19 {\mathrm fm}^{-3},T=10\mbox{
  MeV},P_{\mathrm c.m.}=0)=0
$$
For comparison a perturbative result using Gaussian functions
fitted to the charge radius of the $\alpha$-particle is also given.

\begin{figure}[t]
\begin{center}
\psfig{figure=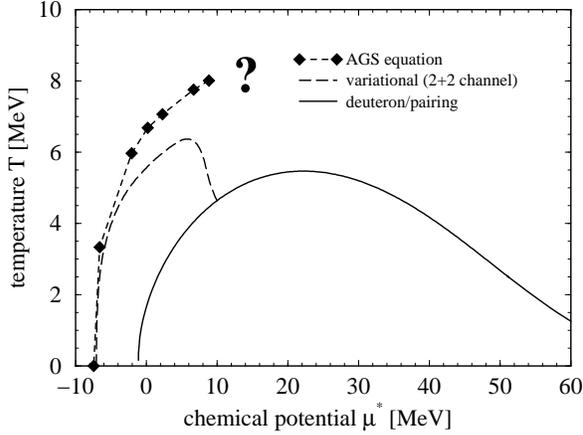,width=0.7\textwidth}
\end{center}
\caption{\label{fig:ProcTc} Temperature vs. effective chemical potential 
  ($\mu^*=\mu-\Sigma(0)$), $\Sigma(0)$ Hartree-Fock shift of the
  single particle energy at zero momentum.  Lines show the critical
  temperature for pairing (solid) and quartetting. The dashed line
  shows a result given in \protect\cite{roe98}, the diamonds show the
  solution of the AGS equation given in (\ref{eqn:coup4})}
\end{figure}
Finally, I address the question of a possible four-particle condensate
or quartetting~\cite{roe98}.  The condition is $B_4(n,T_c,P_{\mathrm
  c.m.}=0)=4\mu $. From Fig.\ref{fig:amottC} we argue that $\alpha$
condensation is likely, i.e.
$$
T_c^{\alpha}>T_c^{\mathrm NN},
$$
where the critical temperature for $\alpha$ condensation turns out
to be higher than for the pairing. However, for $\mu>0$ the situation
seems not so clear, since the four-body AGS equation (\ref{eqn:coup4})
develops poles related to zeros in $4\mu-H_0$. Unlike the two-body
case were these poles disappear because the numerator becomes as well
zero at $2\mu$ and $T_c$ (viz. $1-f_1-f_2\rightarrow 0$). A vanishing
numerator is not obvious for the four-body case because there are more
channels involved. It remains to be clarified, if the rapid fall of
the critical temperature found in Ref.~\cite{roe98} using variational
treatment with square-integrable functions remains, if one uses an
exact treatment of the four-body problem. This is currently
investigated.

{\em Acknowledgment} I am very grateful to S. Mattiello, C. Kuhrts,
G.  R\"opke, P. Schuck, S.A. Sofianos, and W.  Schadow, who have
contributed in certain stages to some results presented here, and to
T. Frederico for lively discussions. I gratefully acknowledge the warm
and pleasant atmosphere at the Department of Physics during stays at
UNISA, Pretoria. Work supported by Deutsche Forschungsgemeinschaft
and University of South Africa.


\begin{references}
\vspace*{-1.5cm}
\bibitem{duk98} J. Dukelsky, G. R\"opke, P. Schuck, Nucl. Phys. A 628,
  17 (1998).  
\bibitem{Beyer:1996rx}
M.~Beyer, G.~Ropke and A.~Sedrakian,
Phys.\ Lett.\ {\bf B376} (1996) 7
[nucl-th/9601038].
\bibitem{Beyer:1997sf}
M.~Beyer and G.~Ropke,
nucl-th/9706021.
\bibitem{Beyer:1999tm}
M.~Beyer,
Few Body Syst.\ Suppl.\ {\bf 10} (1999) 179
[nucl-th/9809002].
\bibitem{Beyer:1999zx}
M.~Beyer, W.~Schadow, C.~Kuhrts and G.~Ropke,
Phys.\ Rev.\ C {\bf 60} (1999) 034004
[nucl-th/9902074].
\bibitem{Kuhrts:2000jz}
C.~Kuhrts, M.~Beyer and G.~Ropke,
Nucl.\ Phys.\ {\bf A668} (2000) 137
[nucl-th/9908032].
\bibitem{Beyer:1999xv}
M.~Beyer, C.~Kuhrts, G.~Ropke and P.~D.~Danielewicz,
nucl-th/9910058.
\bibitem{Beyer:2000ds}
M.~Beyer, S.~A.~Sofianos, C.~Kuhrts, G.~Ropke and P.~Schuck,
Phys.\ Lett.\ {\bf B488} (2000) 247
[nucl-th/0003071].
\bibitem{Kuhrts:2000zs}
C.~Kuhrts, M.~Beyer, P.~Danielewicz and G.~Ropke,
nucl-th/0009037.
\bibitem{alt67} E.O. Alt, P.
  Grassberger, W. Sandhas, Nucl. Phys. {\bf B 2} (1967) 167.
\bibitem{san74} W. Sandhas, Acta Physica Austriaca, Suppl. {\bf XIII}, 
  679 (1974).
\bibitem{alt72} E.O. Alt, P. Grassberger and W. Sandhas, Report
  E4-6688, JINR, Dunba 1972 and in {\em Few particle problems in the
    nuclear interaction} eds. I. Slaus et al. (North Holland,
  Amsterdam 1972) p. 299.
\bibitem{san75} W. Sandhas, Czech. J. Phys. {\bf B 25}, 251 (1975).
\bibitem{fet71} L.P. Kadanoff, G. Baym, {\em Quantum Theory of
    Many-Particle Systems} (Mc Graw-Hill, New York, 1962);
 A.L. Fetter, J.D. Walecka, {\em Quantum Theory of
    Many-Particle Systems}, (McGraw-Hill, New York, 1971).
\bibitem{mat00} S. Mattiello, diploma thesis, 2000
  (unpublished) 
\bibitem{bey01} M. Beyer and S.A. Sofianos, in preparation.
\bibitem{roe98} G. R\"opke, A. Schnell, P. Schuck, P. Nozi\'ere, Phys.
    Rev. Lett. {\bf 80} (1998) 3177
\end{references}
\end{document}